\documentstyle[12pt]{article}
\begin{document}
\begin{center}
{\large \bf 
Multifractal Features in the Foreign Exchange and Stock Markets\\

\vspace*{.5in}

\normalsize 
Kyungsik Kim$^{*}$ and Seong-Min Yoon$^{\dagger}$\\

\vspace*{.2in}

{\em $^{*}$Department of Physics, Pukyong National University, Pusan 608-737, Korea\\
} 
%


\hfill\\
%
%
{\em $^{\dagger}$Division of Economics, Pukyong National University,\\
Pusan 608-737, Korea\\}
}
\end{center}
%
\hfill\\
\baselineskip 24pt
%
\begin{center}
{\bf Abstract}
\end{center} 
   
The multifractal behavior for tick data of prices is investigated in Korean financial market. 
Using the rescaled range analysis(R/S analysis), we show the multifractal nature of returns for 
the won-dollar exchange rate and the KOSPI.
We also estimate the Hurst exponent and the generalized $q$th-order Hurst exponent
in the unversal multifractal framework.
Particularly, our financial market is a persistent process with long-run memory effects,
and the statistical value of the Hurst exponents occurs the crossovers 
at charateristic time scales.
It is found that the probability distribution of returns is well consistent with a Lorentz distribution,
significantly different from fat-tailed properties.\\
\vspace*{.3in}

\noindent
PACS numbers:  02.50.-r, 02.60.-x, 02.70.-c\\
KEYWORDS: Won-dollar exchange rate; KOSPI; Hurst exponent; height-height correlation; R/S analysis
\vskip 3mm
\noindent
$^{*}$E-mail: kskim@pknu.ac.kr

\newpage

\noindent

\hfill\\

In recent years, there has considerable interest in various applications of physical and economic methods
in natural and social sciences $[1]$.   
The representive topics in econophysics have mainly included 
the price changes in open market $[2]$, the distribution of income of companies,
the scaling relation of size fluctuations of companies, the financial analysis of foreign 
exchange rates$[3]$, the tick data analysis of bond futures $[4]$, the herd behavior of financial markets $[5]$,
and the self-organized segregation $[6]$.  
The essential issues with fluctuations have particularly led to a better understanding for
the scaling properties based on methods and approaches in scientific fields.
It was remarkably argued in the previous work $[3]$ 
that the price fluctuations follow the anomalous power law
from the stochastic time evolution equation, which is clearly represented in terms of the Langevin-type equation. 
In view of some studies, the power law distribution, the stretched exponential
distribution, and the fat-tailed distribution have generally elucidated the functional properties
from the numerical results obtained in diverse econophysical models.

On the other hand, in order to measure the multifractals of dynamical dissipative systems, the generalized dimension 
and the spectrum have effectively used to calculate the trajectory of chaotic attractors 
that may be classified by the type and number of the unstable periodic orbits.
Several attempts $[7-9]$ to compute these quantities have primarily been presented from the box-counting method.   
Recently, we have usually used 
the box-counting method to analyze precisely generalized dimensions and scaling exponents for mountain heights 
and sea-bottom depths $[10]$.
For the standard analysis, since there exists notably no statistical correlations between observations,     
the R/S analysis has extended to distinguish the random time series from correlated ones.
The recent work$[11]$ on Norwegian and US stock markets has showed that there exists the
notable persistence caused by long-memory in the time series.

For the volume of bond futures, Scalas $et$ $al.$ $[4]$ have studied the correlation function 
for bond walks from the time series of 
BTP(Buoni del tesoro Poliennali) futures exchanged at 
the London International Financial Futures and options Exchange(LIFFE).  
They have discussed that the continuous-time random walk theory,
formerly introduced by Montroll and Weiss $[12]$, is sucessfully applied to the dynamical behavior
of empirical scaling laws by a set of tick-by-tick data in financial markets.
Mainardi $et$ $al.$ $[13]$ have also dealt with the waiting-time distribution for bond futures traded at 
LIFFE.
The theoretical and numerical arguments for the volume of bond futures traded at 
Korean Futures Exchange market(KOFEX) of Pusan were presented in our previous work $[14]$.
To our knowledge, it is of fundemental importance 
to treat with the multifractal nature of prices for the won-dollar exchange rate and 
the KOSPI(Korean stock price index). The studies of multifractals in Korean financial market
havenot explored up to now. 
The purpose of this letter is to investigate mainly the generic multifractal behavior for tick data of prices 
using the R/S analysis for the won-dollar exchange rate and the KOSPI. 
Particularly, the multifractal Hurst exponents, the height-height correlation 
function, and the probability distribution of returns are also discussed
with long-run memory effects.

To quantify the Hurst exponents, we employ the R/S analysis that is generally contributed 
to estimate the multifractals of a time series $[15,16]$.
First of all, we let consider a price time series of length $n$ given by
$ \lbrace p(t_1 ),  p(t_2 ), ...,p (t_n )  \rbrace$,
and the price $\tau$-returns $r(\tau)$ having time scale $\tau$ and length $n$ that is represented in terms of
$r(\tau) = \lbrace r_1 (\tau),  r_2 (\tau), ...,r_n (\tau) \rbrace$,
with $r_i (\tau)=\ln p(t_i +\tau) - \ln p(t_i )$.
After dividing the time series or returns into $N$ subseries of length $M$, we label each subseries 
$E_{M,d} (\tau )=\lbrace r_{1,d} (\tau ),r_{2,d} (\tau ),..., r_{M,d} (\tau ) \rbrace$, with $d=1,2,...,N $. 
Then, the deviation $ D_{M,d} (\tau )$ 
can be defined directly from the mean of returns ${\bar{r}}_{M,d} (\tau )$
as
\begin{equation}
D_{M,d} (\tau )= \sum_{k=1}^{M}( r_{k,d} (\tau )- {\bar{r}}_{M,d} (\tau )).
\label{eq:f6}
\end{equation} 
The hierachical average value $(R/S)_M (\tau )$ that stands for the rescaled/normalized relation between 
$R_{M,d} (\tau )$ and $S_{M,d} (\tau )$ becomes  
\begin{equation}
(R/S)_M (\tau )= \frac{1}{N} \sum_{d=1}^{N} \frac{R_{M,d} (\tau )}{S_{M,d} (\tau )}
 \propto M^{H(\tau )},
\label{eq:h8}
\end{equation} 
where
the subseries $ R_{M,d} (\tau )$ and the standard deviation $S_{M,d} (\tau )$ are, respectively, given by
%
\begin{eqnarray}
R_{M,d} (\tau ) &=& max \lbrace D_{1,d} (\tau ), D_{2,d} (\tau ),...,D_{M,d} (\tau ) \rbrace \nonumber \\
&&-min  \lbrace D_{1,d} (\tau ), D_{2,d} (\tau ),...,D_{M,d} (\tau ) \rbrace
\label{eq:g7}
\end{eqnarray}
and
\begin{equation}
S_{M,d} (\tau )=
[ \frac{1}{M} \sum_{k=1}^{M}( r_{k,d} (\tau )- {\bar{r}}_{M,d} (\tau ))^2]^{\frac{1}{2}}.
\label{eq:e5}
\end{equation} 
Here $H(\tau )$ is called the Hurst exponent and the reationship between the fractal dimension
$D_f$ and the Hurst exponent $H(\tau )$  can be written as $D_f = 2- H(\tau ) $.
As is well known, we can dynamically evolve the Hurst exponent in the following way: 
(1) The time series is persistent if 
$H(\tau) \in (0.5, 1.0] $. It means that this is characterized by long-run memory affecting all time scales.
One has increasing persistence as $H(\tau)$ approaches $1.0$. 
The persistance process means that the chances will continue to be up or down in the future, if the price 
is up or down.
(2) When $H(\tau)=0.5$, the time series is uncorrelated, and this case is really included to Gaussian or gamma white-noise process. 
The stochastic process with $H(\tau) \neq 0.5$ are also referred to as fractional Brownian motions.
(3) One has antipersistence if $H(\tau) \in (0, 0.5] $.
Hence, it is important to note that the persistent process has little noises, while 
it shows high-frequency noise in the antipersistent process.

To investigate the multifractal properties systematically, 
several methods have been suggested for more than one decade. 
In particular, Barab$\acute{a}$si $et$ $al.$ $[8]$ have recently reported 
the multifractality of self-affine fractals and have 
also studied the multi-affine function and the multifractal spectrum.    
For simplicity, the $q$-th height-height correlation function $F_q (\tau)$ 
that depends only on the time lag $\tau$ takes the form
\begin{equation}
F_q (\tau)=<|p(t+\tau)-p(t)|^q >  \propto {\tau}^{qH_q }, 
\label{eq:j10}
\end{equation} 
where $H_q$ is the generalized $q$th-order Hurst exponent and 
the angular brackets denote a statistical average over time. It would be in reality expected that a nontrivial
multi-affine spectrum can be obtained as $H_q$ varies with $q$.
This has exploited in the multifractal method and the large fluctuation
effects in the dynamical behavior of the price can be explored from Eq.$(5)$.
In our scheme, we will make use of Eqs.$(2)$ and $(5)$ to compute the multifractal features of
prices. The mathematical techniques discussed in this letter lead us to more general results.
%
%

For charactristic analysis of the won-dollar exchange rate and the KOSPI in Korean financial market,
we will present more detailed numerical data of Hurst exponents from the results of R/S analysis.
The generalized $q$th-order Hurst exponents in the height-height correlation function are 
further estimated, and the form of the probability distribution of returns is discussed.
In this letter, 
the tick data for the won-dollar exchange rate were taken from April $1981$ to December $2002$,
while we used the tick data of the KOSPI transacted for $23$ years from April $1981$.
In Fig.$1$ we show the price time series for the the won-dollar exchange rate, in which 
the time step between ticks is evoluted for one day.  
For the measure of the Hurst exponent from Eq.$(2)$, we restrict ourselves to 
three cases of $\tau=$ $1$, $5$, and $24$, although the time interval can be extended to large number 
in our simulation. The Hurst exponents for the won-dollar exchange rate and the KOSPI was obtained numerically 
from the results of R/S analysis, as summarized in Tables $1$ and $2$.
The Hurst exponents for the KOSPI are $H(\tau= 1)$=$0.6886$ and  $H(\tau=30)$=$0.7332$, as plotted in Fig.$2$, and 
it is in fact found that our Hurst values are significantly different from a
well-studied random walk with $H=0.5$. 
These are located in the persistence region similar to those of the crude oil prices $[16]$.
Especially, it may be expected that the Hurst exponent is taken anomalously to be 
near $1$ as  the time series proceeds with long-run memory effects.
Since the crossovers in the function $H(\tau)$ are existed in recent studies,
$H(\tau ) $ from our tick data is similarly found to have the existence of crossovers 
at characteristc time $\tau=9$$(\tau=7$ and $35)$ for the won-dollar exchange rate(the KOSPI).

Next we perform the numerical study of Eq.$(5)$ in order to analyze the generalized $q$th-order Hurst exponents 
in the height-height correlation function $F_q (\tau)$.
Tables $1$ and $2$ include the values of the generalized $q$th-order Hurst exponent $H_q$ in height-height correlation function 
for the won-dollar exchange rate and the KOSPI. 
Especially, the values log$(F_q /q)$ for $q=1,2,...,6$ are plotted in
Fig.$3$ for the the won-dollar exchange rate, and the generalized Hurst exponent is taken to be near $0.65$ as $q\to 1$.
The probability distribution of returns is well consistent with a Lorentz distribution
different from fat-tailed properties, as shown in Figs.$4$ and $5$.

In conclusion, we have presented the multifractal measures from the dynamical behavior 
of prices using the R/S analysis for the won-dollar exchange rate and the KOSPI.
The multifractal Hurst exponents, the generalized $q$th-order Hurst exponent, and 
the form of the probability distribution have discussed with long-run memory effects.
Since our Hurst exponents are larger than $0.5$ through R/S analysis, the time series of prices is
meant to be persistent.
Particularly, it is apparent from our data of the Hurst exponent $H(\tau )$ that
the existence of crossovers is similar to that of other result $[16]$.
Moreover, it is found that the probability distribution for all returns is well consistent with a Lorentz distribution.

Since a plethora of tick data support to carry out the dynamical behavior in our stock and foreign exchange markets,
our analysis would assure that it is in fact able to capture the essential multifractal properties in our present
result. 
In future, we will study extensions of the financial analysis for 
the exchange rates transacted in financial markets.
We expect that our key result will be effectively applied to investigate the other tick data 
in Korean financial markets and compared with other calculations 
transacted in other nations in detail.\\

%
%

%

%
%
%
\vskip 20mm
%
{\bf FIGURE  CAPTIONS}

\vspace {5mm}



\noindent
Fig. $1$. Plot of the tick data for the won-dollar exchange rate, where one time step is the transaction
time evoluted for one day. This continuous tick data were taken from April $1981$ to December $2002$.\\
\vspace {5mm}

\noindent
Fig. $2$. Log-log plot of $R/S(\tau)$ at $\tau=1$(circle) and $24$(triangle) for
the won-dollar exchange rate.\\
\vspace {5mm}

\noindent
Fig. $3$. Plot of The $q$-th height-height correlation function $F_q (\tau)$ of the time interval $\tau$ for 
the KOSPI, where the value of slopes is summarized in Table $2$.\\

\vspace {5mm}

\noindent
Fig. $4$. The probability distribution of returns for the KOSPI. 
the dot line is represented in terms of a Lorentz distribution, 
i.e. $ P(r) =$$  \frac{2b}{\pi} \frac{a}{{r}^2 + a^2} $, 
where $ a= 3.0 \times 10^{-4} $$(5.0 \times 10^{-3} )$ and $ b = 9.4 \times 10^{-5}$$(1.1 \times 10^{-4} )$
for the KOSPI(the  won-dollar exchange rate).
\vspace {5mm}

\noindent
Fig. $5$. The probability distribution of all returns for the won-dollar exchange rate. The dashed and solid lines show
the Gaussian and Lorentz distributions, respectively.  \\

\vspace {5mm}

%
%
{\bf TABLE  CAPTIONS}
\vspace {5mm}

\noindent
Table $1$. Summary of values of the Hurst exponent $H(\tau)$ and
the generalized $q$th-order Hurst exponent $H_q$ for the won-dollar exchange rate.\\
\vspace {3mm}\\
\begin{tabular}{lrr} \hline\hline
            $H(\tau)$             &  $H_q $ \\  \hline                           
            $H(\tau= 1) = 0.6886$ &  $H_1 =$$0.6535$   &  $H_4 =$$0.4307 $      \\
            $H(\tau= 5) = 0.7283$ &  $H_2 =$$0.5614 $  &  $H_5 =$$0.3914 $      \\
            $H(\tau=24) = 0.7332$ &  $H_3 =$$0.4859 $  &  $H_6 =$$0.3629$      \\  \hline\hline
\end{tabular}
\vspace {10mm}\\

\noindent
Table $2$. Summary of values of the Hurst exponent $H(\tau)$ and
the generalized $q$th-order Hurst exponent $H_q$ for the KOSPI.\\ 
\vspace {3mm}\\
\begin{tabular}{lrr} \hline\hline
            $H(\tau)$             &  $H_q $ \\  \hline                           
            $H(\tau= 1) = 0.6238$ &  $H_1 =$$0.7791$    &  $H_4 =$$0.4990 $      \\
            $H(\tau= 5) = 0.6575$ &  $H_2 =$$0.5426 $   &  $H_5 =$$0.4767 $  \\
            $H(\tau=24) = 0.7278$ &  $H_3 =$$0.5215 $   &  $H_6 =$$0.4357$      \\  \hline\hline
\end{tabular}

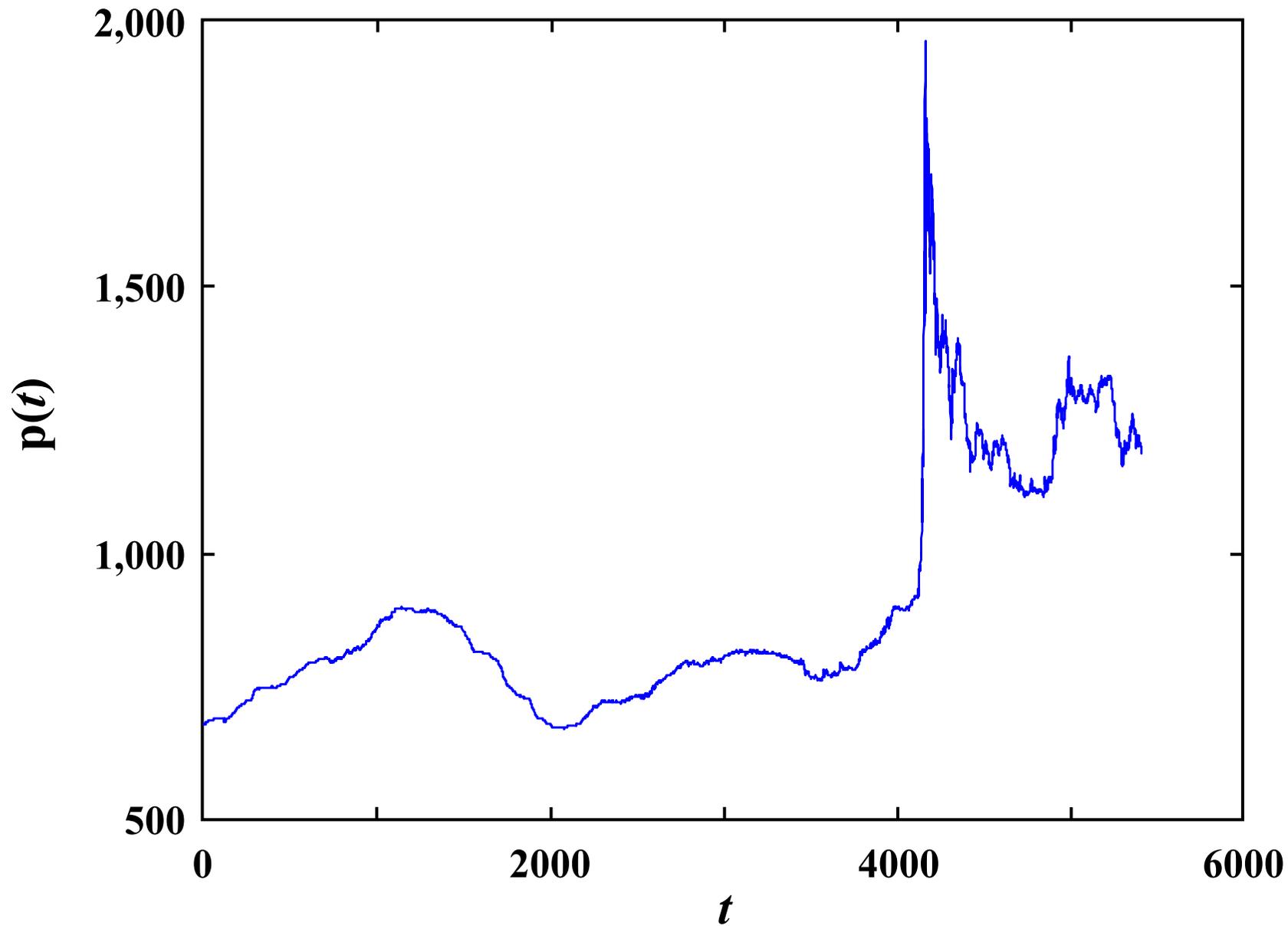
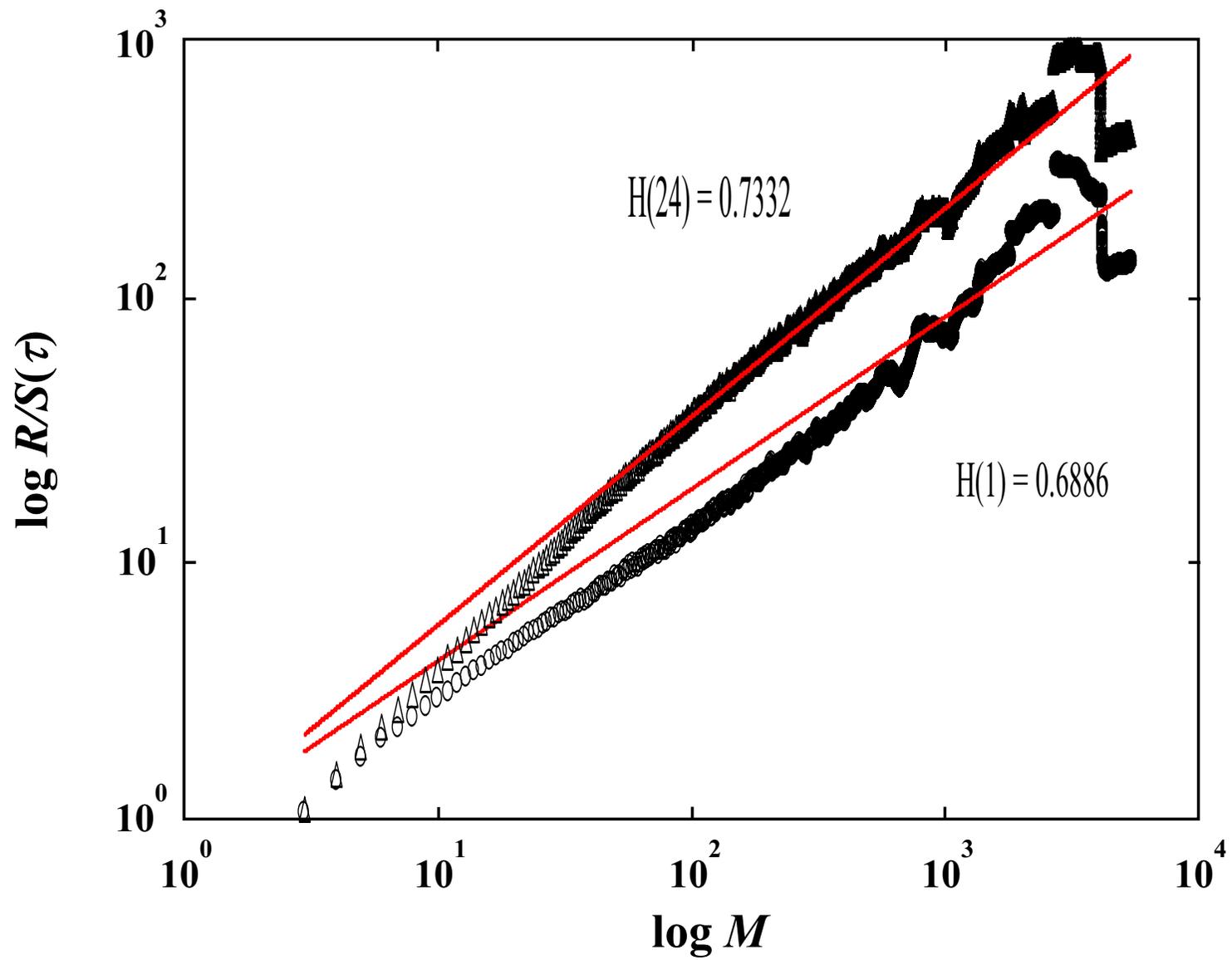
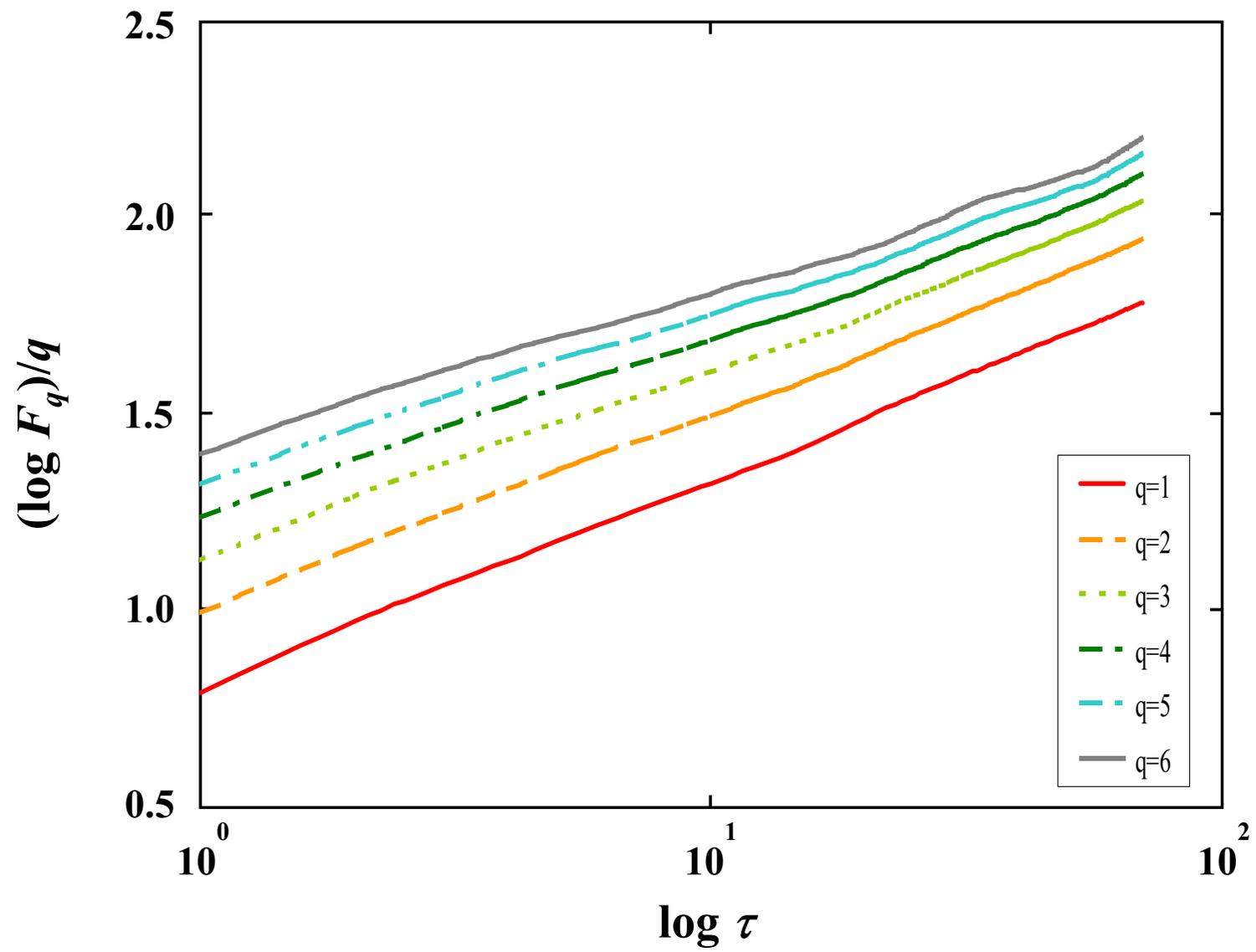
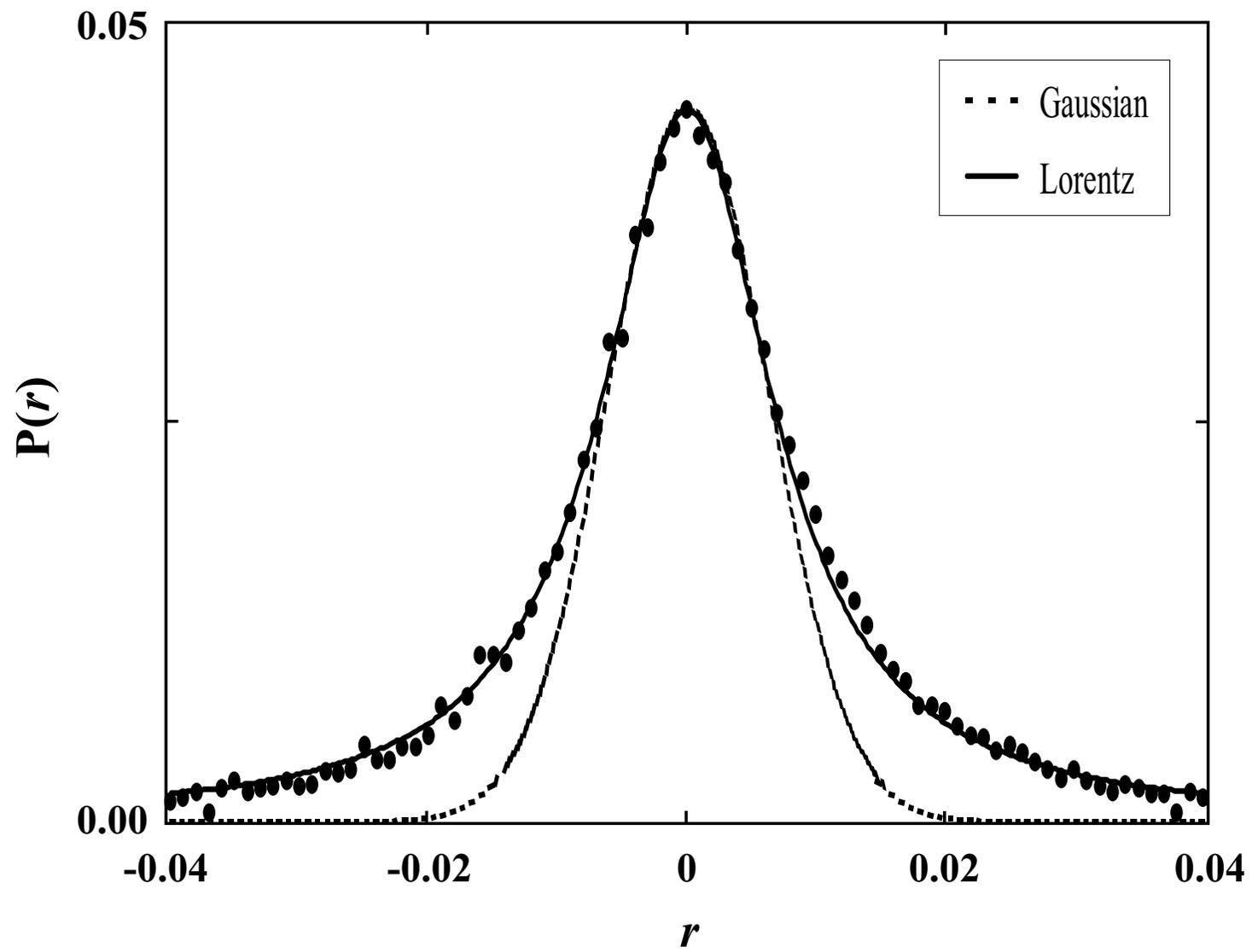
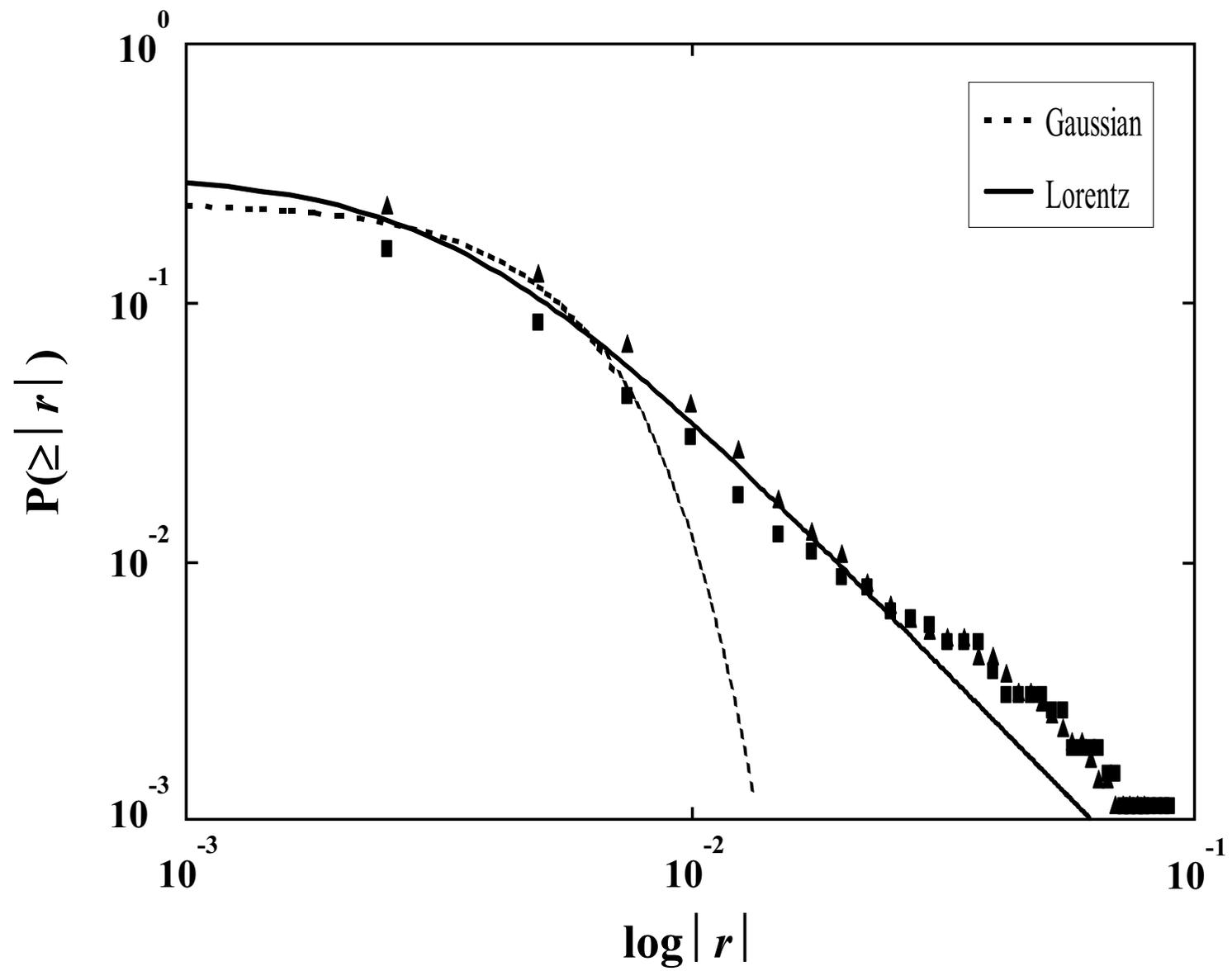
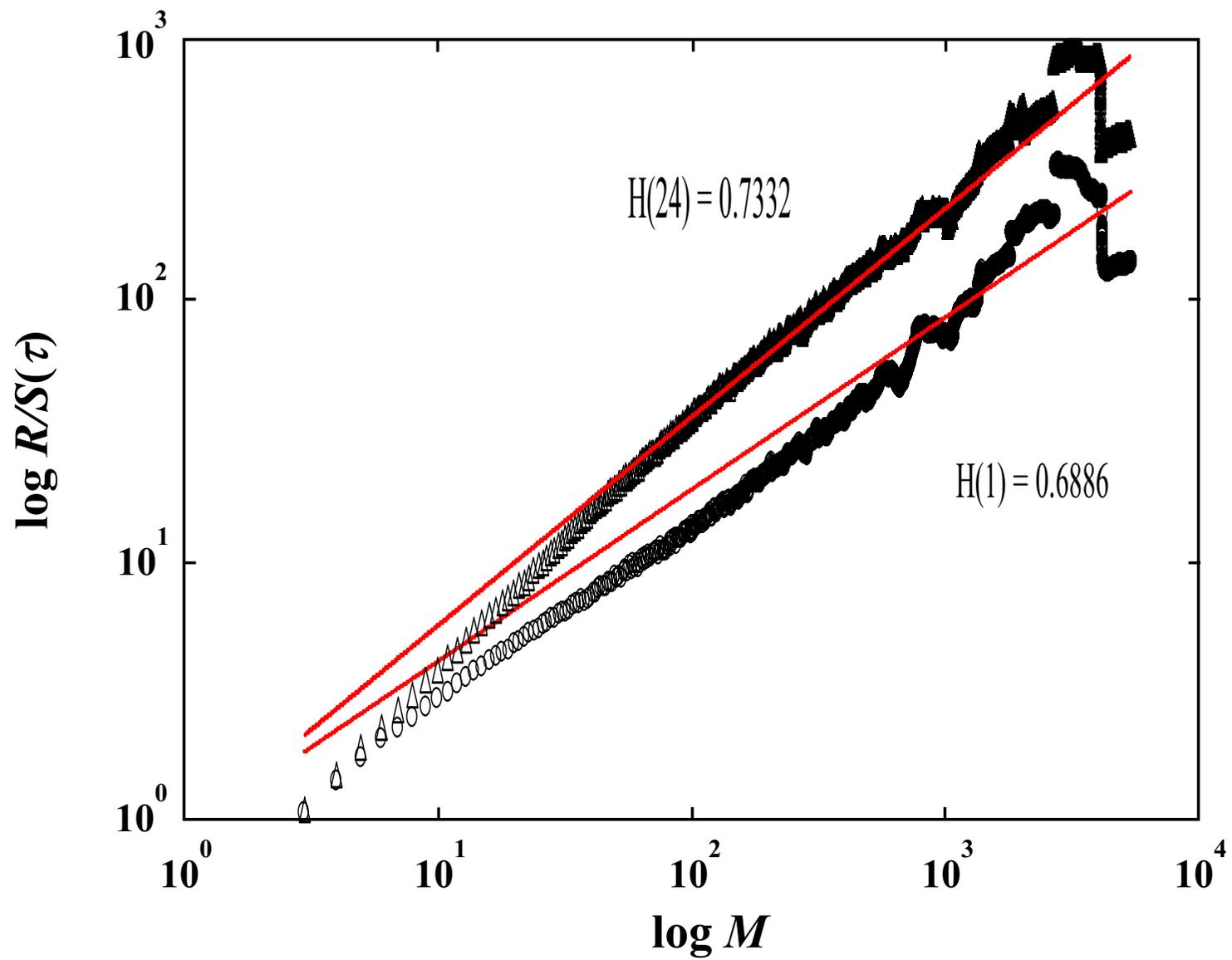
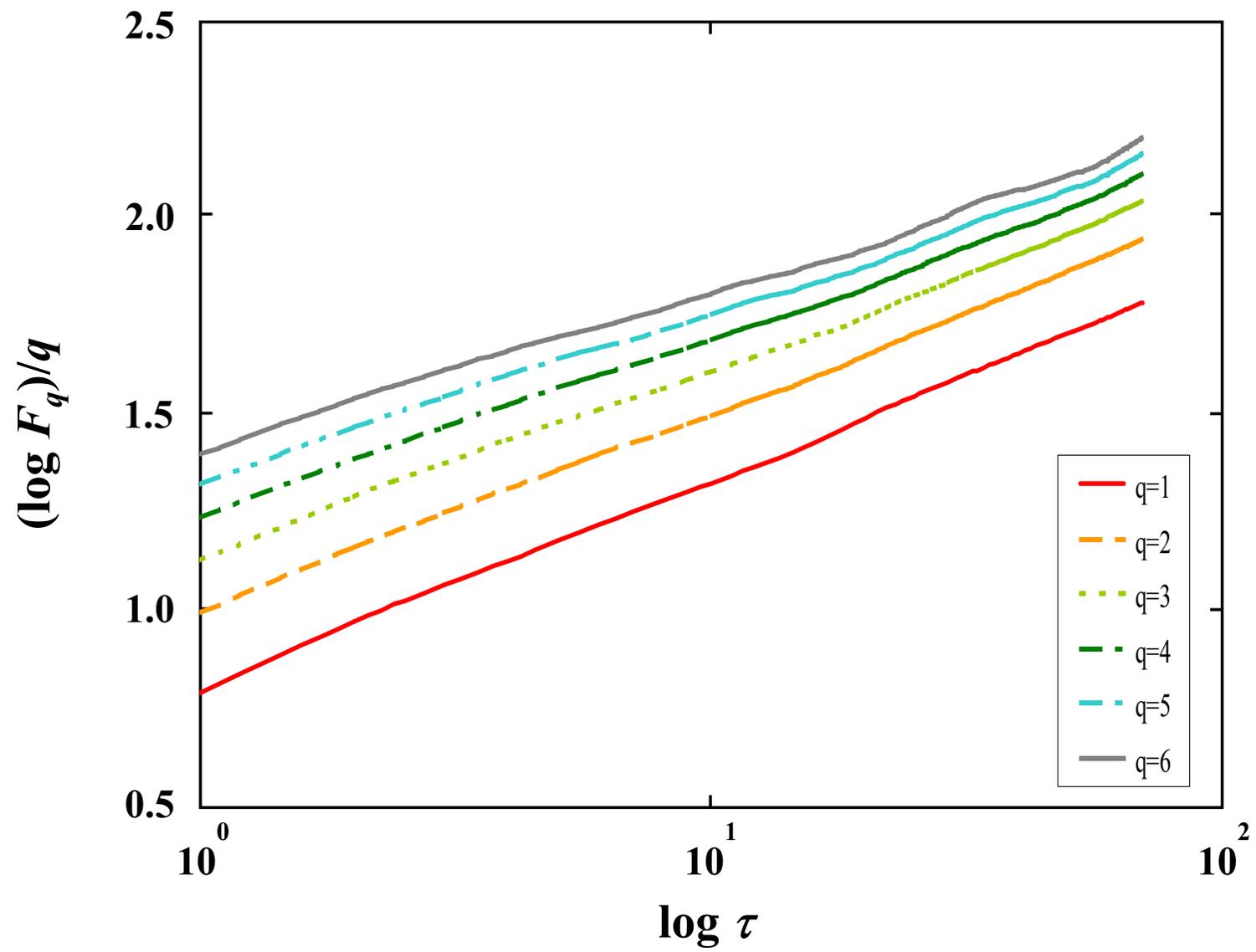
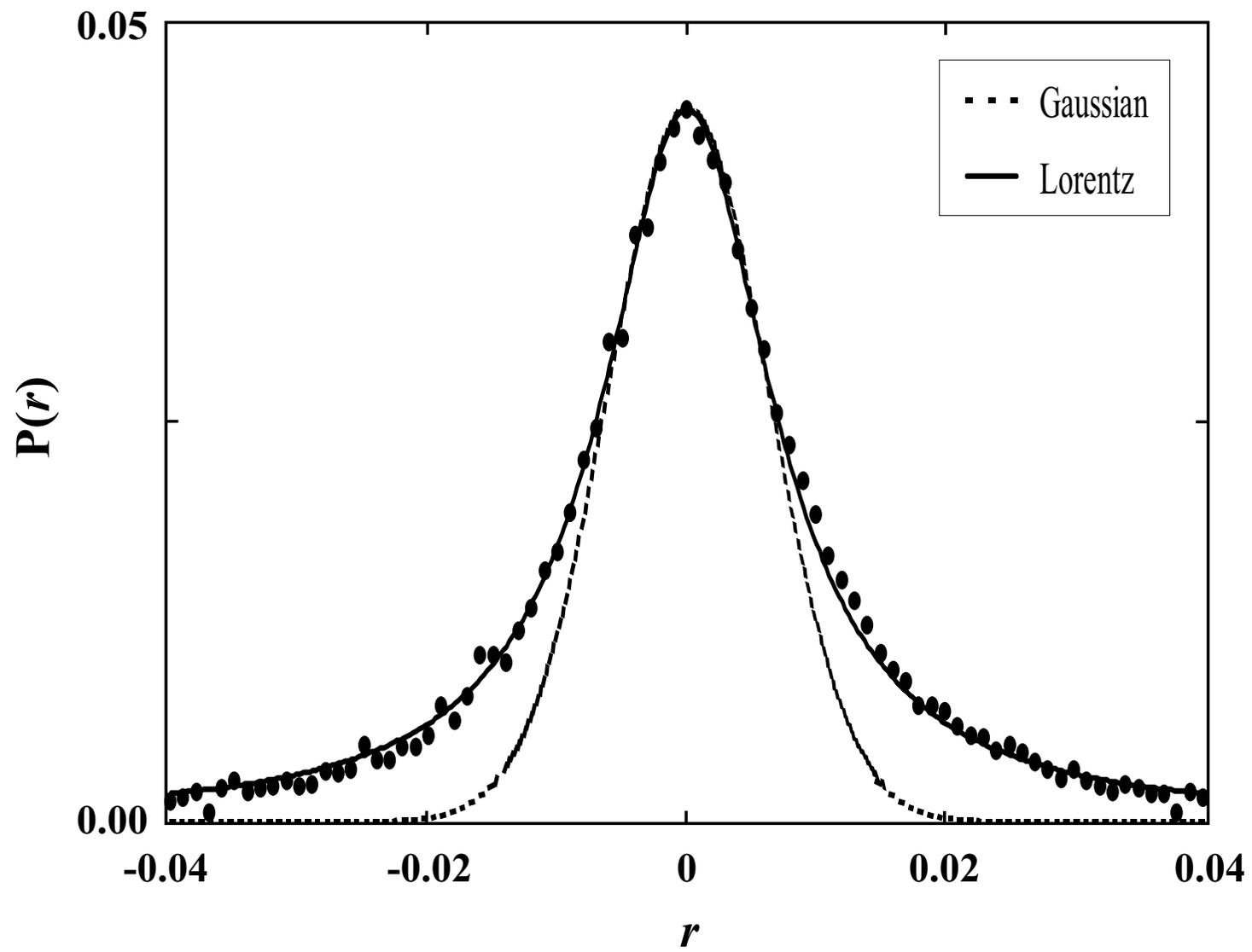
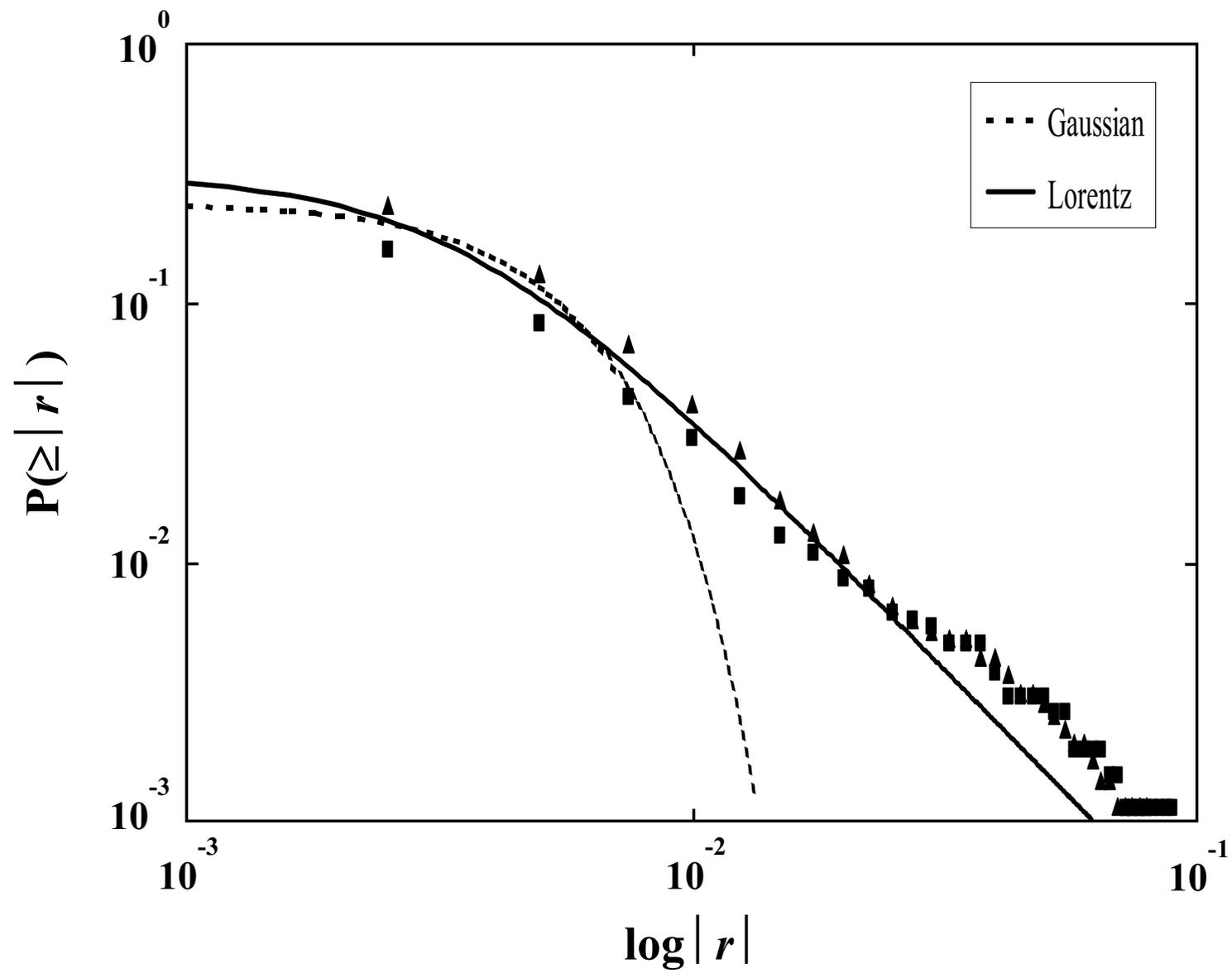
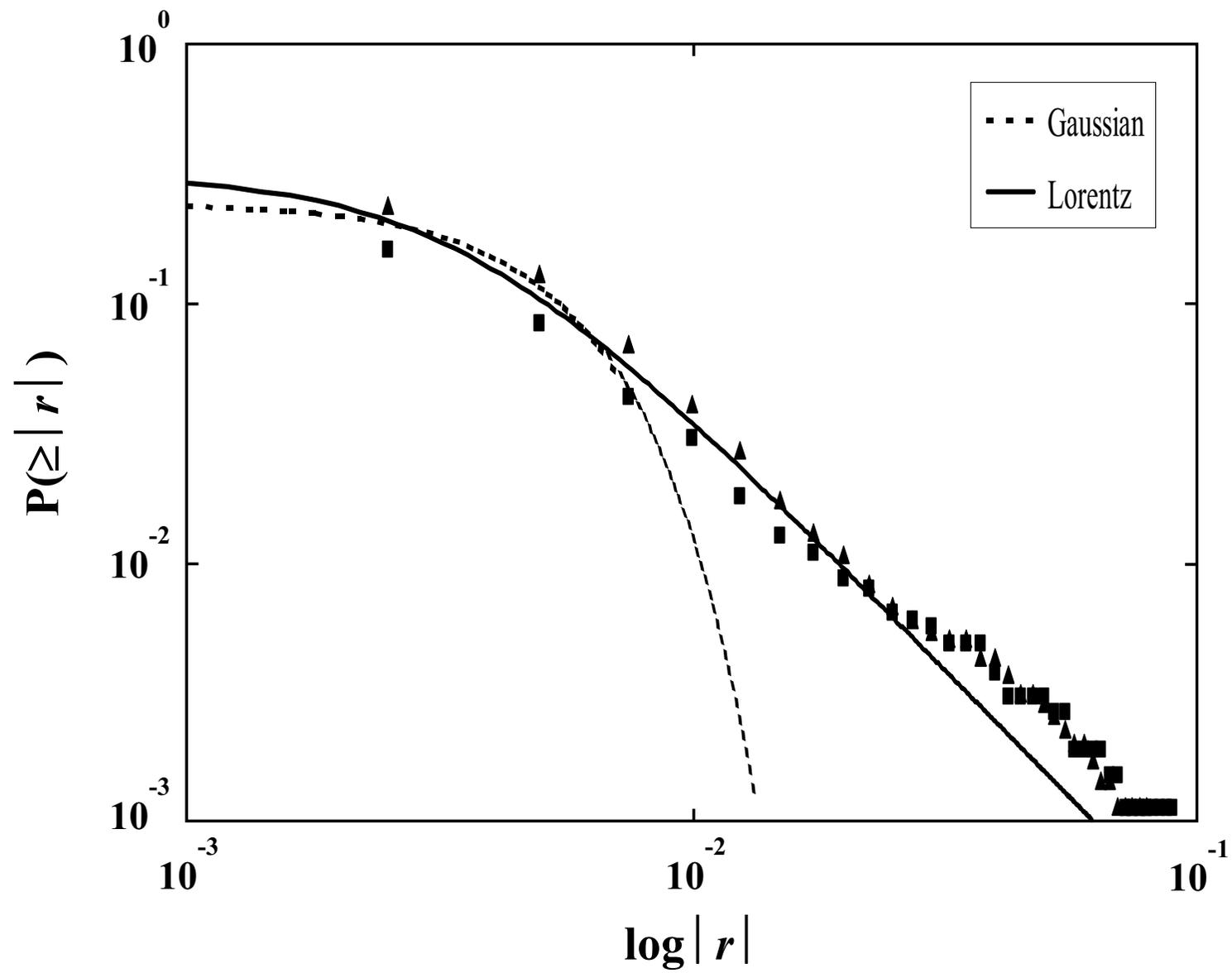
\end{document}